\documentclass{pasj00}

\begin{document}
\SetRunningHead{K. Fujisawa et al.}{Bursting Activity of 6.7~GHz Methanol Maser in G33.64$-$0.21}
\Received{2011/08/06}
\Accepted{2011/09/09}

\title{Bursting Activity in a High-Mass Star-Forming Region G33.64$-$0.21 Observed with the 6.7~GHz Methanol Maser}

\author{%
  Kenta \textsc{Fujisawa},\altaffilmark{1,2}
  Koichiro \textsc{Sugiyama},\altaffilmark{1}
  Nozomu \textsc{Aoki},\altaffilmark{1}
  Tomoya \textsc{Hirota},\altaffilmark{3}
  Nanako \textsc{Mochizuki},\altaffilmark{4}
  Akihiro \textsc{Doi},\altaffilmark{4}
  Mareki \textsc{Honma},\altaffilmark{3}
  Hideyuki \textsc{Kobayashi},\altaffilmark{3}
  Noriyuki \textsc{Kawaguchi},\altaffilmark{3}
  Hideo \textsc{Ogawa},\altaffilmark{5}
  Toshihiro \textsc{Omodaka},\altaffilmark{6}
  and
  Yoshinori \textsc{Yonekura}\altaffilmark{7}}
\altaffiltext{1}{Department of Physics, Faculty of Science, Yamaguchi University, Yoshida 1677-1, Yamaguchi-city, Yamaguchi 753-8512}
\altaffiltext{2}{The Research Institute of Time Studies, Yamaguchi University, Yoshida 1677-1, Yamaguchi-city, Yamaguchi 753-8511}
\altaffiltext{3}{Mizusawa VLBI Observatory, National Astronomical Observatory of Japan, Hoshigaoka-cho 2-12, Oshu, Iwate 023-0861}
\altaffiltext{4}{Institute of Space and Astronautical Science, Japan Aerospace Exploration Agency, Yoshinodai 3-1-1, Chuo-ku, Sagamihara, Kanagawa 252-5210}
\altaffiltext{5}{Graduate School of Science, Osaka Prefecture University, 1-1 Gakuen-cho, Nakaku, Sakai, Osaka 599-8531}
\altaffiltext{6}{Department of Physics, Faculty of Science, Kagoshima University, Korimoto 1-21-24, Kagoshima-city, Kagoshima 890-8580}
\altaffiltext{7}{Center for Astronomy, Ibaraki University, 2-1-1 Bunkyo, Mito, Ibaraki 310-8512}
\email{kenta@yamaguchi-u.ac.jp}

%

\KeyWords{ISM: H\emissiontype{II} regions --- ISM: individual (G33.64$-$0.21) --- masers: methanol} 

\maketitle

\begin{abstract}
We report the detection of bursts of 6.7~GHz methanol maser emission in a high-mass star-forming region, G33.64$-$0.21.
One of the spectral components of the maser in this source changed its flux density by 7 times that of the previous day,
and it decayed with a timescale of 5 days. 
The burst occurred repeatedly in the spectral component, and no other components showed such variability. 
A VLBI observation with the Japanese VLBI Network (JVN) showed that the burst location was at the southwest edge of a spatial distribution, 
and the bursting phenomenon occurred in a region much smaller than 70~AU. 
We suggest an impulsive energy release like a stellar flare as a possible 
mechanism for the burst.
These results imply that 6.7~GHz methanol masers could be a useful new probe for studying bursting activity in the process of 
star formation of high-mass YSOs with a high-resolution of AU scale.
\end{abstract}

\section{Introduction}
A rapid rise of flux called a flare or a burst is often observed in young stellar objects (YSOs). 
An X-ray flare with high temperature gas (e.g., Skinner et al. 1997; Tsuboi et al. 1998) 
and a radio flare with cyclotron or gyro-synchrotron radiation (e.g., Phillips et al. 1996; Massi et al. 2006)
have been observed in V773 Tau, 
which is a binary T Tauri (pre-main sequence) star. 
The timescale of these flares is short, on the order of approximately 0.1 to 10 days. 
Such flares are thought to be caused by magnetic reconnection and its energy release around the star (Bower et al. 2003). 
Because the interaction of the gas and the magnetic field at the stellar surface and the circumstellar environment (disk)
is important in the process of star formation, 
it is necessary to reveal the nature of these flares, i.e. magnetic energy release, with AU scale spatial resolution.
Flare activities have been reported mainly for low-mass YSOs so far. 
Although some authors (Tsuboi et al. 1999; Hamaguchi et al. 2005; Stelzer et al. 2005)
discussed that some part of X-ray variability were magnetic origin, 
high-mass YSOs are not considered active in terms of bursting variability. 

We studied the 6.7~GHz methanol maser which helped us understand not only the gas dynamics
but also the possible flaring activity around high-mass YSOs. 
The methanol masers at 6.7~GHz are observed only around high-mass star-forming regions (Minier et al. 2003; Xu et al. 2008). 
These masers are thought to be in the regions of a gas density of $10^{5-8}$~cm$^{-3}$
and a dust (infrared excitation source) temperature of $>$100~K, 
corresponding to a distance of $\leq $1000~AU from a central star (Cragg et al. 2005). 
Because the size of a maser spot is as compact as several AU (Minier et al. 2002; Sugiyama et al. 2008a), 
the maser can be observed using a Very Long Baseline Interferometer (VLBI). 
The spatial distribution and the radial velocity structure of the maser spots observed by VLBI
often show that a disk or a rotating toroid often exists around high-mass YSOs
(Minier et al. 2000; Bartkiewicz et al. 2009; Moscadelli et al. 2007), 
and the large part of the methanol maser emanates from the gas rotating around or infalling to these YSOs. 
The flux variability of this maser has been observed with long-term monitoring observations, 
and some types of variation, such as periodicity, have been found (Goedhart et al. 2003, 2004, 2009). 
In some sources observed by flux monitoring and spatial distribution with VLBI, 
it is proposed that the variability of the maser is caused by the variation of the exciting source
(Sugiyama et al. 2008b; Araya et al. 2010; Szymczak et al. 2011). 

When a stellar flare occurs, the released energy would heat the gas
and dust around the flare region, causing a maser flare. 
The maser flare can be used to reveal the location at which magnetic
reconnection and/or other activity occurs around the YSO by conducting
a VLBI observation. 
To look for such maser flares, we selected 22 maser sources that have
relatively large flux densities and multiple spectral components, and
carried out a daily monitoring observation. 
The detail of the whole observation will be reported elsewhere,
we report here focusing on the discovery of a bursting methanol maser
in G33.64$-$0.21 (IRAS 18509+0027, \timeform{18h53m32s.563}
 \timeform{+00D31'39''.18}, J2000.0 after Bartkiewicz et al. 2009).

This source is a high-mass star-forming region located at the near-kinematic distance of 4.0~kpc, 
and the infrared luminosity obtained from the IRAS database was $1.2\times 10^4~\LO$. 
Although there are some infrared sources detected with 2MASS, Sptzer/GLIMPSE and MSX near the methanol position,
it is not clear which one is the associating source due to the large uncertainty of the maser position ($\sim$ 5'').
The methanol maser of this source was first observed by the Torun survey (Szymczak et al. 2000). 
There are at least five spectral features (components I-V, defined in radial velocity order)
with a typical line width of 0.3~km~s$^{-1}$ and one weak, broad feature (component VI) in the spectrum (figure 1). 
The magnetic field of this source was measured through polarization observation as $-$18~mG,
which is stable over the entire velocity range (Vlemmings 2008).

In this paper, we describe the spectra monitoring and VLBI observations and the results in section 2,
then we discuss interpretations of the mechanism of the bursting methanol maser in section 3.

\section{Observations and Results}
\subsection{Spectra Monitoring}
The monitoring observations were made with the Yamaguchi 32~m radio telescope
from July 2 (Day-of-Year 183) to October 17 (DOY 290), 2009. 
A total of 62 observations were made over the 108-day observation period. 
The observations were usually made every one or two days, and the maximum interval was five days. 
The spectrometer used for this observation was the IP-VLBI system (Kondo et al. 2003). 
The sampling was made at a rate of 8~Msamples~sec$^{-1}$ (bandwidth 4~MHz) and at two bits per sample. 
The spectra were obtained by a software spectrometer with  16384 spectral channels, then binned into 4096, yielding a velocity resolution of 0.044~km~s$^{-1}$. 
The integration time was 14 minutes. 
The left and right circularly polarized waves were observed simultaneously
and were averaged after being transformed into the spectra. 
A 1$\sigma $ rms noise level is 1.2~Jy. The accuracy of the absolute value of the flux density is approximately 10\%, 
while the standard deviation of the relative fluctuation is 3\% estimated from the daily data.
The $V_\mathrm{LSR}$ calculation may contain $<$0.3~km~s$^{-1}$ constant offset.

\begin{figure}
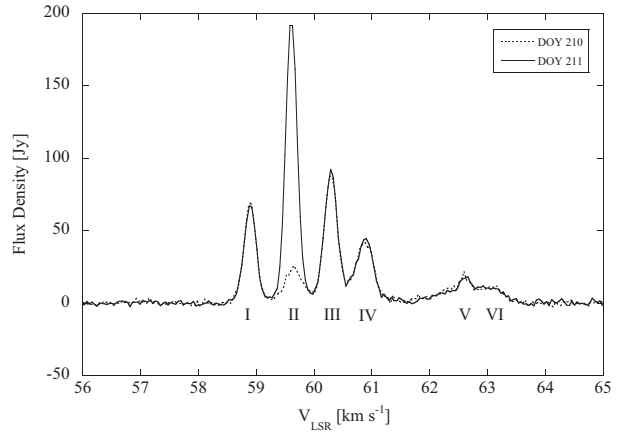

  \begin{center}
    \FigureFile(80mm,80mm){figure1.eps}
  \end{center}
  \caption{Spectra of G33.64$-$0.21. The spectra of G33.64$-$0.21 observed with the Yamaguchi 32~m radio telescope are shown. 
The dotted and solid lines show the spectra of day-of-year (DOY) 210 and 211, respectively, just before and after the second burst. 
Only spectral component II ($V_\mathrm{LSR}=59.6$~km~s$^{-1}$) rose by a factor of seven (26 to 191~Jy).
}\label{fig:fig1}
\end{figure}

Two rapid rises in the flux density (hereafter, burst) were detected in G33.64$-$0.21 at DOY 186 and 211 only
in component II ($V_\mathrm{LSR}=59.6$~km~s$^{-1}$), as shown in figure 2. 
At the first burst, the flux density rose to 7 times within three days (DOY 183 to 186),
and it fell exponentially with an e-folding time of 5 days. 
Because there was three days lack of observation just before the burst, the rising timescale and the peak flux density contain some ambiguity. 
In the second burst, the flux density rose from 26~Jy to 191~Jy, i.e., a factor of 7 times
within 24 hours from DOY 210 to 211, and then it fell slowly. 
The decrement of the flux density was exponential, with an e-folding time of 5 days. 
This behavior is similar to that of the first burst. A small burst was observed at DOY 262. 
There was no change whatsoever in the other spectral components when the bursts occurred. 
The flux density of component I decreased from 75 to 20~Jy from DOY 208 to DOY 240. 
Such variation with a timescale of 20 days has been observed in other sources
(Sugiyama et al. 2008b; Goedhart et al. 2009).

\begin{figure}
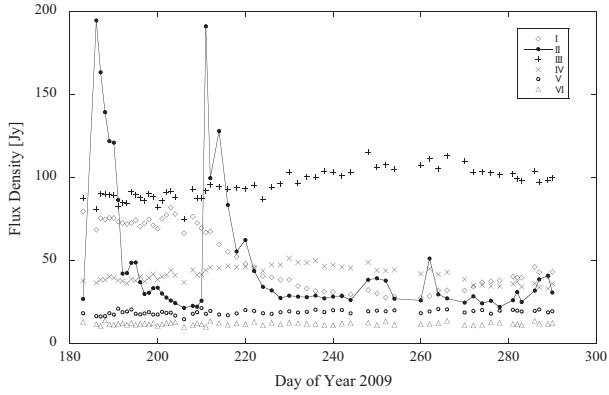

  \begin{center}
    \FigureFile(80mm,80mm){figure2.eps}
  \end{center}
  \caption{The variation of spectral components. The changes of peak flux density for each six spectral components (I-VI) are shown. 
Only component II (filled circle and the solid line) shows the burst, while the other components are completely unrelated to the burst. 
Two large bursts (DOY 186, 211) and a small burst on DOY 262 were seen during the 108 days of observation run.
}\label{fig:fig2}
\end{figure}

The peak velocity of component II was measured by gaussian fitting. The velocity
slightly decreased (0.05~km~s$^{-1}$) when the component bursted, 
and the velocity returned to its original value after the burst (figure 3). 
This velocity change repeated for the two bursts with an accuracy of less than 0.01~km~s$^{-1}$. 
This velocity shift may be explained as blending two spectral components with almost the same but slightly different velocities; 
one component has high velocity and stable flux, and the other has low velocity and shows the bursting variability. 
This small velocity variation and its repeatability suggest that the bursting region
is not disturbed by violent gas motion like a shockwave even when they showed large bursts. 

Szymczak et al. (2010) reported that G33.64$-$0.21 showed a large flare only in the spectral component II
(59.6~km~s$^{-1}$) near JD=2455160 (DOY 330 of 2009).
The observed ratio of the maximum and minimum of the flux density was 16.7.
This suggests that G33.64$-$0.21 bursted again at this time.
They also reported the decay time of $\sim 30$ days,
however, this might be an wrong estimation due to their sparse observation (intervals of 1 to 6 weeks).

A bursting phenomenon like the one observed in G33.64$-$0.21 was found only in this source among the 22 sources observed. 
A large bursting variability in methanol masers was reported for NGC6334F (Goedhart et al. 2004). 
The variation pattern of NGC~6334F, i.e., the fast rise and relatively slow fall, looks like the burst of G33.64$-$0.21; 
however, the timescale of the burst in NGC6334F is as long as 100 days.
Periodic flare phenomena have been found in some sources of methanol maser.
Araya et al. (2010) reported a periodic flare of formaldehyde and methanol masers.
The duration of the flare is $\sim 30$ days for IRAS~18566$+$0408, while it is only 5 days for G33.64$-$0.21.
Szymczak et al. (2011) discussed the time profile of the flares of two sources; G22.357$+$0.066 and G9.62$+$0.20E
(the data is from Goedhart et al. 2004).
The rise and fall (decay) time are $14.5 \pm 1.7$ and $42.9 \pm 0.7$ days for G22.357$+$0.066.
These are ten times larger than those of G33.64$-$0.21.
The ratios of the rise and fall time are 0.34 and 0.37 for G22.357$+$0.066 and G9.62$+$0.20E, respectively,
while it is less than 0.2 for G33.64$-$0.21. The time asymmetry is much obvious for G33.64$-$0.21.

\begin{figure}
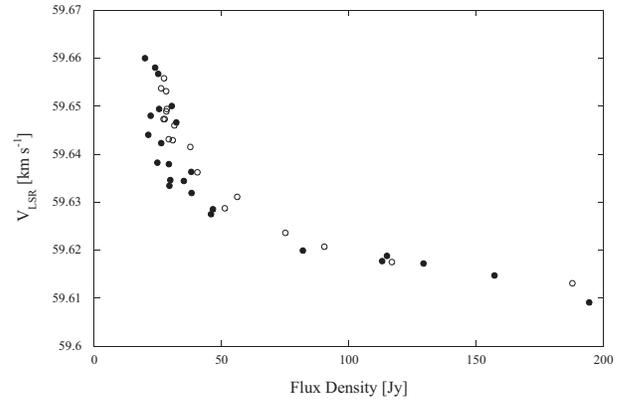

  \begin{center}
    \FigureFile(80mm,80mm){figure3.eps}
  \end{center}
  \caption{Peak velocity and flux density of component II. The change of peak velocity of
component II with respect to its flux density is shown.
A negative correlation is obvious.
Filled circles are for the first burst (DOY 183 to 210),
and open circles are for the second burst (DOY 211 to 248), respectively.
The velocity changes repeated for the two bursts with an accuracy of less than 0.01~km~s$^{-1}$.
}\label{fig:fig3}
\end{figure}

\subsection{VLBI Observation}
A VLBI observation using the Japanese VLBI Network (JVN) was conducted on October 6, 2009 (DOY 279) from 05:00 to 14:00 UT,
to determine the position in G33.64$-$0.21 where the burst occurred.
Although the burst did not occur at this time, the bursted component (II) was visible.
Five JVN telescopes (Yamaguchi 32~m and four VERA 20~m telescopes) were used.
The observation frequency was 6665-6669~MHz, and the number of frequency channels was 1024.
The spatial distribution of the 35 maser spots is shown in figure 4.
Those maser spots were detected with a signal-to-noise ratio larger than 5, and at least 2 adjacent frequency channels.
The synthesized beam size was 11.7$\times $3.0~mas$^{2}$ with a position angle of $-$45.2~deg, as shown in the lower-left corner of the figure.
The rms noise of the image in a line-free channel is 70~mJy~beam$^{-1}$.

\begin{figure}
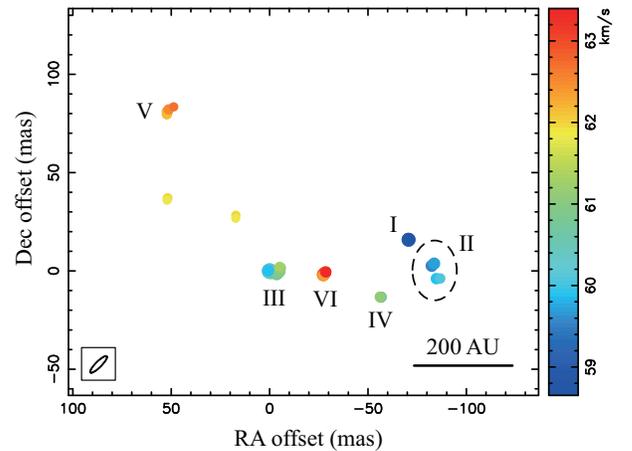

  \begin{center}
    \FigureFile(80mm,80mm){figure4.eps}
  \end{center}
  \caption{Distribution of maser spots observed with the JVN. 
The spot size and color indicate its peak intensity in logarithmic scale and its radial velocity (see color index at the right), respectively.
The origin of this figure is the position of the strongest component at $V_\mathrm{LSR}=60.23$~km~s$^{-1}$.
The maser spots are distributed in an arched structure with an extension of 650$\times $200~AU.
Labels of I to VI correspond to the spectral components.
The maser spots of the bursting spectral component (II) are located at the southwest edge of the distribution, as indicated with the dashed ellipse.
}\label{fig:fig4}
\end{figure}

The maser spots form an arched structure with a size of 160~mas, corresponding to the linear size of 650~AU at 4.0~kpc. 
This distribution is essentially the same as the result by Bartkiewicz et al. (2009). 
There is no peculiar characteristic in the spatial distribution of G33.64$-$0.21
compared with those of other sources shown by VLBI surveys (Minier et al. 2000; Dodson et al. 2004; Sugiyama et al. 2008a; Bartkiewicz et al. 2009). 
Component II, which showed the burst, is located at the southwest edge of the spatial distribution of the maser spots. 
There are two clusters of spots that are 30~AU apart in this region comprising spectral component II: it is unknown which of them caused the burst.
The VLBI image reconstructed the flux density of 62\% of the Yamaguchi spectrum for the component II.
This suggest that the size of the component II is as small as the beam size (11.7$\times $3.0~mas$^{2}$).
The spots of spectral components I and IV are located 70~AU in northeast and 130~AU in east direction away from the clusters of component II. 
Because spectral components I and IV did not show any change when component II bursted, 
the burst-related phenomenon occurred within a spatial scale of much less than 70~AU.

\section{Discussions}
The observed properties of the burst are summarized as follows: (1) the rising timescale was less than one day, whereas the falling was 5 days; 
(2) the burst only occurred in spectral component II, at the southwest edge of the spatial distribution;
and (3) the bursting phenomenon occurred in a region much smaller than 70~AU. 
Before concluding that the magnetic energy release was the plausible origin of the burst, other possibilities are considered.

It was reported that interstellar scintillation caused the variability of a hydroxyl maser (Clegg \& Cordes 1991). 
The amplitude (7 times) and time scale (less than 1 day) of the burst in G33.64$-$0.21 are, however, the strong arguments against that the scintillation is
the origin of the burst. In addition, if
the scintillation was the origin of the burst, all spectral components would show some variability;
the bursts were observed only in one spectral component of G33.64$-$0.21.
Therefore, the burst was not caused by interstellar scintillation.

In some maser sources, the flux variability of the methanol maser is proposed to be caused
by the variability of the luminosity of the exciting source
(Sugiyama et al. 2008b; Araya et al. 2010; Szymczak et al. 2011). 
The variation pattern of the flux density should be synchronized or correlated between
two or more spectral components if the maser variation is caused by the luminosity variation of the exciting source. 
In the case of G33.64$-$0.21, however, not even the spectral component nor maser spots
spatially adjacent to the bursting spots within a distance of only 70~AU varied when component II bursted. 
Therefore, the burst of G33.64$-$0.21 was not caused by a variation of the exciting source.

Shimoikura et al. (2005) showed by VLBI monitoring that the burst of the 22~GHz water
maser in Orion KL in 1998 was caused by an overlapping and amplifying effect that occurred between two maser clouds, 
which were on the line of sight by chance. 
If we apply the overlapping model to the burst of G33.64$-$0.21, its short timescale would be a difficulty. 
Because the spectrum of G33.64$-$0.21 has a velocity range of 6~km~s$^{-1}$,
one can assume 3~km~s$^{-1}$, half of the range, to be the speed of the gas cloud. 
The maser cloud can move only $3\times 10^8$~m in one day at such speed, corresponding to 0.5~$\mu $as at a distance of 4.0~kpc. 
The brightness temperature would exceed 10$^{18}$~K if a flux rise of 165~Jy occurred in this small region. 
This brightness temperature cannot be achieved with the radiation mechanism of the 6.7~GHz methanol maser. 
Moreover, the asymmetry of the variation of rise (timescales of less than 1 day) and fall (5 days) cannot be explained with this model.

The simplest explanation of this burst is that energy was injected within a short time into the small region at which the burst occurred. 
An example of such energy release seen in astrophysics is the solar flare, which is an impulsive event of magnetic field reconnection. 
The energy accumulated in the magnetic field on the solar surface is discharging by the reconnection of the magnetic field in a short time. 
The released energy causes a particle acceleration that heats the gas near the reconnection position, and finally, radiation is emanated from the gas. 
Similar to the solar flare, flares at a much larger scale are observed in T Tauri stars,
such as V773 Tau (Phillips et al. 1996) or GMR-A (Bower et al. 2003; Furuya et al. 2003). 
Here, we propose a mechanism of the maser burst: an impulsive energy release which resembles those observed in active YSOs
occurred at the stellar surface or in the circumstellar disk around the high-mass YSO in G33.64$-$0.21, causing the maser burst. 
The released energy was converted into the heat of the gas and dust surrounding the release point. 
The population inversion of the methanol molecule was strengthened by the infrared emission from the heated dust. 
Finally, the maser emission emanating from the gas around the release point was suddenly strengthened. 
The asymmetry of the rise and fall of the burst is naturally explained within this model
because the rise is impulsive, but the fall depends on the relatively slow radiative cooling process.
The cooling time scale was theoretically estimated of 1.2 days by van der Walt et al. (2009) for the optically thick condition.
This time scale was found to be too short to explain the decay time of NGC6334F maser flare ($\sim 100$ days),
but could be applicable to the case of G33.64$-$0.21, which has significantly shorter decay time ($\sim 5$ day).
Because the velocity change during the burst was small, the shock process is not suitable to explain this burst. 
A theoretical calculation (Cragg et al. 2005) would support our scenario because the brightness
of the maser depends on the dust temperature, while the gas density and gas temperature are insensitive parameters
 to the brightness temperature variation.
A change of dust temperature of 100~K (e.g. 150 to 250~K) would cause the increment of maser flux of 7 times.

A brief consideration of energy dynamics is presented below. 
Assuming the size of the area where the gas is heated is 10~AU in diameter
and the number density of the hydrogen molecule is 10$^6$~cm$^{-3}$, then the gas mass in this area would be $6\times 10^{21}$~kg. 
An energy of $6\times 10^{27}$~J is required to heat the gas and raise
its temperature by 100~K so that the maser excitation condition changes drastically. 
This energy is comparable to the energy of an X-ray flare observed in V773 Tau (10$^{27}$~J; Tsuboi et al. 1998). 
If we assume that the energy was released from the magnetic field in/around the star, and that
the magnetic field strength was 18~mG throughout this region, the total magnetic energy in this volume would be $2\times 10^{30}$~J. 
Because the magnetic energy is much higher than the required energy for the burst, the magnetic energy release can account for the burst.

The large bursts occurred twice, and they only occurred in spectral component II,
which means that the energy releases repeatedly occurred at the southwest edge of G33.64$-$0.21. 
At the bursting place, there might be a binary companion or planet,
which could cause a local disturbance and trigger the energy release. 
The distribution of a water maser of G33.64$-$0.21 (Bartkiewicz et al. 2011)
is elongated north and south from the position near the bursting methanol component (II), 
but the positional relation between the water maser and the central star
is not definitive because of the large positional uncertainties of 5$^{\prime \prime }$ and 0.15$^{\prime \prime }$
of the methanol and water maser, respectively.

If the maser burst was caused by local heating, a local burst would be observed simultaneously at infrared and submillimetre bands. 
Such observation requires high spatial resolution, which will be achieved by ALMA (Atacama Large Millimeter/submillimeter Array). 
Finally, we emphasize that this result demonstrates the 6.7~GHz methanol
maser can be a useful tool to investigate the bursting activity in the process of star formation around high-mass YSOs.

\section{Conclusion}
We have presented the discovery of the bursting activity in a high-mass star-forming region detected with the 6.7~GHz methanol maser. 
A 108-day continuous monitoring for 22 sources of methanol maser was made with the Yamaguchi 32~m radio telescope. 
In one source, G33.64$-$0.21, rapid rises of the flux density (burst) were observed twice only in one spectral component out of 6 components. 
In the second burst, the flux density rose 7 times within 24 hours, and then it fell exponentially with an e-folding time of 5 days. 
There was absolutely no change in the other spectral components when the bursts occurred. 
A VLBI observation using the Japanese VLBI Network (JVN) showed that
the maser spots form an arched structure and the burst component is located at the southwest edge of the spatial distribution of the maser spots. 
There are spots which did not show burst 70~AU from the burst spot. 
This means that the burst-related phenomenon occurred within a spatial scale of much less than 70~AU. 
In order to account these properties, we proposed a mechanism of the maser burst as
an impulsive energy release like the stellar flare observed in YSOs
occurred at the stellar surface or in the circumstellar disk in G33.64$-$0.21. 
Finally, we noted that the 6.7~GHz methanol maser can be a useful tool
to investigate the bursting activity in the process of star formation around high-mass YSOs.

\bigskip

The authors thank the JVN team for observation assistance and support. 
The JVN project is led by the National Astronomical Observatory of Japan (NAOJ),
which is a branch of the National Institutes of Natural Sciences (NINS), 
Hokkaido University, Ibaraki University, Tsukuba University, Gifu University,
Osaka Prefecture University, Yamaguchi University, and Kagoshima University, 
in cooperation with the Geographical Survey Institute (GSI),
the Japan Aerospace Exploration Agency (JAXA),
and the National Institute of Information and Communications Technology (NICT). 
K.F. thanks Professor Mashiyama for his encouragement of this study, the KDDI corporation
for supporting the Yamaguchi 32~m radio telescope, and Mr. Nagadomi for assisting with data analysis. 
This work was supported by a Grant-in-Aid for Scientific Research (B) (20340045).



\begin{thebibliography}{}
\bibitem[Araya et al.(2010)]{2010ApJ...717L.133A} 
Araya, E.~D., Hofner, P., Goss, W.~M., Kurtz, S., Richards, A.~M.~S., Linz, H., Olmi, L., \& Sewi{\l}o, M.\ 2010, \apjl, 717, L133 
\bibitem[Bartkiewicz et al.(2009)]{2009A&A...502..155B}
Bartkiewicz, A., Szymczak, M., van Langevelde, H.~J., Richards, A.~M.~S., \& Pihlstr{\"o}m, Y.~M.\ 2009, \aap, 502, 155
\bibitem[Bartkiewicz et al.(2011)]{2011A&A...525A.120B} 
Bartkiewicz, A., Szymczak, M., Pihlstr{\"o}m, Y.~M., van Langevelde, H.~J., Brunthaler, A., \& Reid, M.~J.\ 2011, \aap, 525, A120 
\bibitem[Bower et al.(2003)]{2003ApJ...598.1140B} 
Bower, G.~C., Plambeck, R.~L., Bolatto, A., McCrady, N., Graham, J.~R., de Pater, I., Liu, M.~C., \& Baganoff, F.~K.\ 2003, \apj, 598, 1140 
\bibitem[Clegg \& Cordes(1991)]{1991ApJ...374..150C} 
Clegg, A.~W., \& Cordes, J.~M.\ 1991, \apj, 374, 150 
\bibitem[Cragg et al.(2005)]{2005MNRAS.360..533C} 
Cragg, D.~M., Sobolev, A.~M., \& Godfrey, P.~D.\ 2005, \mnras, 360, 533 
\bibitem[Dodson et~al.(2004)]{2004MNRAS.351..779D} 
Dodson, R., Ojha, R., \& Ellingsen, S.~P.\ 2004, \mnras, 351, 779 
\bibitem[Furuya et al.(2003)]{2003PASJ...55L..83F} 
Furuya, R.~S., Shinnaga, H., Nakanishi, K., Momose, M., \& Saito, M.\ 2003, \pasj, 55, L83 
\bibitem[Goedhart et al.(2003)]{2003MNRAS.339L..33G} 
Goedhart, S., Gaylard, M.~J., \& van der Walt, D.~J.\ 2003, \mnras, 339, 33 
\bibitem[Goedhart et al.(2004)]{2004MNRAS.355..553G} 
Goedhart, S., Gaylard, M.~J., \& van der Walt, D.~J.\ 2004, \mnras, 355, 553 
\bibitem[Goedhart et al.(2009)]{2009MNRAS.398..995G}
Goedhart, S., Langa, M.~C., Gaylard, M.~J., \& van der Walt, D.~J.\ 2009, \mnras, 398, 995 
\bibitem[Hamaguchi et al.(2005)]{2005ApJ...618..360H}
Hamaguchi, K., Yamauchi, S., \& Koyama, K.\ 2005, \apj, 618, 360
\bibitem[Kondo et al.(2003)]{2003ASPC..306..205K} 
Kondo, T., Koyama, Y., Nakajima, J., Sekido, M., \& Osaki, H.\ 2003, Astronomical Society of the Pacific Conference Series, 306, 205 
\bibitem[Massi et al.(2006)]{2006A&A...453..959M} 
Massi, M., Forbrich, J., Menten, K.~M., Torricelli-Ciamponi, G., Neidh{\"o}fer, J., Leurini, S., \& Bertoldi, F.\ 2006, \aap, 453, 959 
\bibitem[Minier et~al.(2000)]{2000A&A...362.1093M}
Minier, V., Booth, R.~S., \& Conway, J.~E.\ 2000, \aap, 362, 1093 
\bibitem[Minier et al.(2002)]{2002A&A...383..614M} 
Minier, V., Booth, R.~S., \& Conway, J.~E.\ 2002, \aap, 383, 614 
\bibitem[Minier et~al.(2003)]{2003A&A...403.1095M}
Minier, V., Ellingsen, S.~P., Norris, R.~P., \& Booth, R.~S.\ 2003, \aap, 403, 1095 
\bibitem[Moscadelli et~al.(2007)]{2007A&A...472..867M}
Moscadelli, L., Goddi, C., Cesaroni, R., Beltr{\'a}n, M.~T., \& Furuya, R.~S.\ 2007, \aap, 472, 867 
\bibitem[Phillips et al.(1996)]{1996AJ....111..918P} 
Phillips, R.~B., Lonsdale, C.~J., Feigelson, E.~D., \& Deeney, B.~D.\ 1996, \aj, 111, 918 
\bibitem[Shimoikura et al.(2005)]{2005ApJ...634..459S} 
Shimoikura, T., Kobayashi, H., Omodaka, T., Diamond, P.~J., Matveyenko, L.~I., \& Fujisawa, K.\ 2005, \apj, 634, 459 
\bibitem[Skinner et al.(1997)]{1997ApJ...486..886S} 
Skinner, S.~L., Guedel, M., Koyama, K., \& Yamauchi, S.\ 1997, \apj, 486, 886 
\bibitem[Stelzer et al.(2005)]{2005ApJS..160..557S}
Stelzer, B., Flaccomio, E., Montmerle, T., Micela, G., Sciortino, S., Favata, F., Preibisch, T., \& Feigelson, E. D. \ 2005, \apjs, 160, 557
\bibitem[Sugiyama et al.(2008a)]{2008PASJ...60...23S} 
Sugiyama, K., Fujisawa, K., Doi, A., Honma, M., Kobayashi, H., Bushimata, T., Mochizuki, N., \& Murata, Y.
\ 2008a, \pasj, 60, 23 
\bibitem[Sugiyama et al.(2008b)]{2008PASJ...60.1001S}
Sugiyama, K., Fujisawa, K., Doi, A., Honma, M., Isono, Y., Kobayashi, H., Mochizuki, N., \& Murata, Y.\ 2008b, \pasj, 60, 1001
\bibitem[Szymczak et al.(2000)]{2000A&AS..143..269S} 
Szymczak, M., Hrynek, G., \& Kus, A.~J.\ 2000, \aaps, 143, 269 
\bibitem[Szymczak et al.(2010)]{2010evn..confE..99S} 
Szymczak, M., Bartkiewicz, A., Wolak, P. Proceedings of the 10th European VLBI Network Symposium and EVN Users Meeting: VLBI and the new generation of radio arrays. September 20-24, 2010. Manchester, UK.
\bibitem[Szymczak et al.(2011)]{2011A&A...531L...3S}
Szymczak, M., Wolak, P., Bartkiewicz, A., \& van Langevelde, H.~J.\ 2011, \aap, 531, L3 @
\bibitem[Tsuboi et al.(1998)]{1998ApJ...503..894T} 
Tsuboi, Y., Koyama, K., Murakami, H., Hayashi, M., Skinner, S., \& Ueno, S.\ 1998, \apj, 503, 894 
\bibitem[Tsuboi et al.(1999)]{1999AN....320..175T} 
Tsuboi, Y., Hamaguchi, K., Koyama, K., \& Yamauchi, S.\ 1999, Astronomische Nachrichten, 320, 175 
\bibitem[van der Walt et al.(2009)]{2009MNRAS.398..961V} 
van der Walt, D.~J., Goedhart, S., Gaylard, M.~J.\ 2009, \mnras, 398, 961 
\bibitem[Vlemmings(2008)]{2008A&A...484..773V} 
Vlemmings, W.~H.~T.\ 2008, \aap, 484, 773 
\bibitem[Xu et~al.(2008)]{2008A&A...485..729X}
Xu, Y., Li, J.~J., Hachisuka, K., Pandian, J.~D., Menten, K.~M., \& Henkel, C.\ 2008, \aap, 485, 729 
\end{thebibliography}
\end{document}